\documentclass[letter]{jpsj3}

\title{Microcanonical Analysis of Spin Glasses Using Gauge Symmetry}

\author{Hidetoshi Nishimori} 

\inst{Department of Physics, Tokyo Institute of Technology,\\
Meguro-ku, Oh-okayama, Tokyo 152-8551, Japan
}
\abst{We apply the method of gauge transformation to spin glasses under the
microcanonical ensemble to study the possibility of ensemble inequivalence
in systems with long-range interactions and quenched disorder.
It is proved that all the results derived under the canonical ensemble on the Nishimori line (NL)
can be reproduced by the microcanonical ensemble irrespective of the range of interactions.
This establishes that ensemble inequivalence should take place away from the NL
if it happens in spin glasses.
It is also proved on the NL that the microcanonical configurational average of
the energy as a function of temperature is exactly equal to the average energy in the
canonical ensemble for any finite-size systems with Gaussian disorder.
In this sense, ensembles are equivalent even for finite systems.
}

\kword{microcanonical ensemble, ensemble inequivalence, spin glass, gauge symmetry}

\begin{document}
\maketitle

Equivalence of canonical and microcanonical ensembles is well
established for systems with short-range interactions.\cite{LL,Greiner}
It is not necessarily the case in the presence of long-range
interactions because of the absence of additivity:\cite{Campa}
If two independent systems with long-range interactions are put together,
the total energy is not equal to the sum of the energies of two separate
systems, and thus the standard derivation of the canonical ensemble
from the microcanonical ensemble breaks down.
One of the prominent consequences is the emergence of negative specific heat
in the microcanonical ensemble, which has long been discussed in the astrophysical
context \cite{Ant,Lynden,Thirring} and has been observed in condensed matter
of small size where the range of interactions is comparable to the system size.\cite{Cn}
Ensemble inequivalence has also been discussed in spin systems with long-range interactions
with \cite{BKMN} and without \cite{spin1,spin2,spin3,spin4} quenched disorder.
It is therefore important to establish exact/rigorous results on ensemble equivalence
and inequivalence that are applicable generically to a class of problems.
The present contribution represents a step toward this goal.

More specifically, we develop a general theory of spin glasses using gauge symmetry
in the microcanonical ensemble, which should be compared with the corresponding
theoretical framework in the conventional canonical ensemble,
in particular on the so-called Nishimori line (NL).\cite{PTP81,Book2001}
The results agree with those from the canonical ensemble, whenever comparison
is possible, irrespective of the range of interactions including
the infinite-range case.
It is therefore concluded that canonical and microcanonical ensembles
are equivalent in spin glasses on the NL as long as physical
quantities that can be analyzed by gauge symmetry are concerned.

Let us study the Edwards-Anderson model of spin glasses,\cite{EA}
\begin{equation}
 H=-\sum_{\langle ij\rangle} J_{ij} S_i S_j,\label{Hamiltonian}
\end{equation}
where the $J_{ij}$ are quenched random interactions and the $S_i(=\pm 1)$ are Ising spins.
The range of interactions is arbitrary and includes the infinite-range
Sherrington-Kirkpatrick model.\cite{SK}
The lattice structure and spatial dimensionality are also arbitrary.
The present theory is thus very general.

We first discuss the Gaussian distribution of $J_{ij}$ implemented
in the microcanonical context.
Our microcanonical analysis starts from the entropy for a given energy, averaged over disorder,
\begin{equation}
S(E)=\frac{1}{C_J}\int {\rm D}J\, \delta \big(N_BJ_0-\sum_{\langle ij\rangle} J_{ij}\big)
\log \sum_S \delta (E-H),
\label{S}
\end{equation}
where $C_J$ is the normalization factor of the configurational average
\begin{equation}
C_J=\int {\rm D}J\, \delta \big(N_BJ_0-\sum_{\langle ij\rangle} J_{ij}\big)
\end{equation}
with $N_B$ being the number of bonds, and ${\rm D}J$ is the Gaussian integral kernel
\begin{equation}
{\rm D}J=\prod_{\langle ij\rangle}\frac{{\rm d}J_{ij}}{\sqrt{2\pi}}\exp(-J_{ij}^2/2).
\end{equation}
In eq. (\ref{S}), the configurational average is taken under the condition
$\sum J_{ij}=N_B J_0$.
This is the microcanonical version of the Gaussian distribution with average $J_0$,
which corresponds to the standard (canonical) weight,
\begin{equation}
\exp \Big( -\frac{1}{2}\sum_{\langle ij\rangle} (J_{ij}-J_0)^2\Big)
=\exp \Big(-\frac{1}{2}\sum_{\langle ij\rangle}(J_{ij}^2+J_0^2)\Big)
\exp \Big( J_0\sum_{\langle ij\rangle} J_{ij}\Big).
\label{Gauss}
\end{equation}
The final factor in the above equation indicates that the value of
the sum $\sum J_{ij}$ can fluctuate according to the probability
weight $\exp(J_0 \sum J_{ij})$.
The microcanonical version of eq. (\ref{S}) 
strictly enforces the value of $\sum J_{ij}$ to $N_B J_0$.

It is convenient to evaluate here the normalization factor $C_J$ using the
condition $E=-N_B J_0$ that will be explained later,
\begin{eqnarray}
C_J(E)&=&\frac{1}{2\pi}\int {\rm d}\lambda\, \exp(-i\lambda E)
\int {\rm D}J\,\exp(-i\lambda \sum J_{ij}) \nonumber\\
&=&\frac{1}{2\pi}\int {\rm d}\lambda\, \exp(-i\lambda E-N_B\lambda^2/2)
=\frac{1}{\sqrt{2\pi N_B}}\exp\big(-E^2/(2N_B)\big).
\label{C}
\end{eqnarray}

Let us apply the gauge transformation $J_{ij}\to J_{ij}\sigma_i \sigma_j,~S_i\to S_i \sigma_i$
($\forall i, j$), where the $\sigma_i=\pm 1$ are gauge variables.
Following the prescription established in the canonical ensemble,\cite{PTP81,Book2001}
we sum the result over all configurations of the gauge variables to have
\begin{equation}
S(E)=\frac{1}{2^NC_J}\int {\rm D}J\,\sum_\sigma \delta 
\big(N_BJ_0-\sum_{\langle ij\rangle} J_{ij}\sigma_i\sigma_j\big)
\log \sum_S \delta \big(E+\sum_{\langle ij\rangle} J_{ij}S_i S_j)\big),
\end{equation}
where $N$ is the number of sites.
The basic strategy of microcanonical calculations is to take the derivative of
the entropy with respect to the energy and equate the result to the inverse temperature $\beta$.
If we carefully look at the expression of this derivative,
\begin{equation}
\beta =\frac{\partial S}{\partial E}=\frac{1}{2^NC_J}\int {\rm D}J\,\sum_\sigma \delta 
\big(N_BJ_0-\sum_{\langle ij\rangle} J_{ij}\sigma_i\sigma_j\big)
\frac{\partial_E\sum_S \delta \big(E+\sum_{\langle ij\rangle} J_{ij}S_i S_j)\big)}
{\sum_S \delta \big(E+\sum_{\langle ij\rangle} J_{ij}S_i S_j)\big)},
\end{equation}
we notice that the summation over $\sigma$ in the configurational average
cancels with the summation over $S$ in the denominator when $E=-N_B J_0$.
This is the microcanonical condition for the NL.
Then the above expression reduces to
\begin{eqnarray}
\beta &=&\frac{\partial S}{\partial E}=\frac{1}{2^NC_J(E)}\partial_E \sum_S \int {\rm D}J\,
\delta \big(E+\sum_{\langle ij\rangle} J_{ij}S_i S_j\big) \nonumber\\
&=&\frac{1}{C_J(E)}\partial_E 
\int {\rm D}J\,\delta \big(E+\sum_{\langle ij\rangle} J_{ij}\big)\nonumber\\
&=&\frac{\partial_E C_J(E)}{C_J(E)}=-\frac{E}{N_B},
\label{beta_Gaussian}
\end{eqnarray}
where we have used eq. (\ref{C}).
The second line has been derived by the gauge transformation $J_{ij}\to J_{ij}S_i S_j$.
Thus we have
\begin{equation}
E=-N_B \beta \label{Gaussian_energy}
\end{equation}
in precise agreement with the canonical case.\cite{PTP81,Book2001}
The NL condition in the canonical ensemble $\beta =J_0$ also follows from the condition $E=-N_BJ_0$
in conjunction with eq. (\ref{Gaussian_energy}).
It is remarkable that both ensembles give the same exact energy as a function
of temperature for any finite-size system, any dimension, and any range of interactions.
Usually, the results of two ensembles agree only in the thermodynamic limit.

Upper and lower bounds for the specific heat $C$ for $E=-N_BJ_0$ can also be estimated
as shown below, and the results agree with the corresponding canonical inequalities,
\begin{equation}
0\le C\le \frac{N_B}{T^2}. \label{C-ineq}
\end{equation}
Since the specific heat is non-negative and upper-bounded by the same expression
as in the canonical ensemble, there exists no ensemble inequivalence under the NL condition
at least as long as the energy and bounds on the specific heat are concerned.

To prove eq. (\ref{C-ineq}), we evaluate the second derivative of the entropy
with respect to the energy, using the notation $\Omega (E)=\sum_S\delta (E-H)$,
\begin{eqnarray}
&&\frac{\partial^2 S}{\partial E^2} =\frac{1}{2^N C_J}\int {\rm D}J\, 
\sum_\sigma \delta \big(N_BJ_0-\sum_{\langle ij\rangle} J_{ij}\sigma_i\sigma_j\big) 
\left\{\frac{\partial^2_E \Omega (E)}{\Omega (E)}
-\left(\frac{\partial_E\Omega (E)}{\Omega (E)}\right)^2 \right\}
\nonumber\\
&\le & \frac{1}{2^N C_J}\int {\rm D}J\, \sum_\sigma 
\delta \big(N_BJ_0-\sum_{\langle ij\rangle} J_{ij}\sigma_i\sigma_j\big)
\frac{\partial^2_E\Omega (E)}{\Omega (E)}
\nonumber\\
&&
-\left(
\frac{1}{2^N C_J}\int {\rm D}J\, \sum_{\sigma}\delta \big(N_BJ_0-\sum_{\langle ij\rangle}
 J_{ij}\sigma_i\sigma_j\big)
\frac{\partial_E\Omega (E)}{\Omega (E)}
\right)^2
\nonumber\\
&&=\frac{\partial_E^2 C_J(E)}{C_J(E)}-
\left(\frac{\partial_E C_J(E)}{C_J(E)} \right)^2=-\frac{1}{N_B}.
\end{eqnarray}
If we remember the relation
\begin{equation}
-\frac{1}{CT^2}=\frac{\partial \beta}{\partial E},
\end{equation}
it follows
\begin{equation}
-\frac{1}{CT^2}\le -\frac{1}{N_B},
\end{equation}
which is the desired inequality (\ref{C-ineq}).

The same analysis applies to the $\pm J$ model, where $J_{ij}=1$ with probability $p$
and $J_{ij}=-1$ with probability $1-p$.
We have set $J(=|J_{ij}|)=1$ for simplicity without losing generality.
The points to be modified from the Gaussian case are
(i) to replace the integral over $J_{ij}$ with the summation over $J_{ij}=\pm 1$,
(ii) to replace the normalization $C_J$ with $C_p$, where
\begin{equation}
C_p=\sum_{\{J_{ij}\}} \delta \big(N_B(2p-1)-\sum_{\langle ij\rangle} J_{ij}\big),
\end{equation}
and (iii) to replace the condition $E=-N_BJ_0$ with $E=-N_B(2p-1)$.
The resulting relation corresponding to eq. (\ref{beta_Gaussian}) is
\begin{equation}
\beta=\frac{\partial_E C_p(E)}{C_p(E)}, \label{beta_p}
\end{equation}
where
\begin{equation}
C_p(E)=\int {\rm d}\mu\, \exp(\mu E+N_B\log 2 \cosh \mu).
\end{equation}
Here we have omitted the factor $1/2\pi$ since it plays no role in the following.
The integral variable $\mu$ corresponds to $i\lambda$ in eq. (\ref{C}).
For finite $N_B$, it is impossible to evaluate the integral explicitly,
which is a difference from the Gaussian case.
In the thermodynamic limit, the saddle-point method yields
$\beta =\mu_0$ from eq. (\ref{beta_p}), where $\mu_0$ is the saddle point specified by
\begin{equation}
E+N_B\tanh \mu_0=0.
\end{equation}
In combination with $\beta =\mu_0$ and $E=-N_B(2p-1)$, we conclude $E=-N_B\tanh \beta$
under the condition $\tanh \beta =2p-1$.
This is in perfect agreement with the canonical analysis.\cite{PTP81,Book2001}
Complete ensemble equivalence holds only in the thermodynamic limit in the $\pm J$ model.

Upper and lower bounds for the specific heat on the NL
can also be estimated as in the Gaussian model.
The central inequality is
\begin{equation}
\frac{\partial \beta}{\partial E}\le \frac{\partial_E^2 C_p(E)}{C_p(E)}
-\left(\frac{\partial_E C_p(E)}{C_p(E)}\right)^2. \label{C_pm}
\end{equation}
If we naively apply the saddle-point method and take only the leading term,
the right-hand side reduces $\mu_0^2-\mu_0^2=0$, leading to
\begin{equation}
-\frac{1}{CT^2}\le 0
\end{equation}
or $C\ge 0$.
This positivity of the specific heat is non-trivial in the microcanonical ensemble
but is not very exciting.
A better inequality is obtained from the leading correction to the saddle point.
The normalization factor $C_p(E)$ is written as, to the leading correction to the
saddle point, with the notation $f(\mu)=\mu E+N_B\log 2\cosh \mu$,
\begin{equation}
C_p(E)=\exp\big(f(\mu_0)\big)\int {\rm d}\mu\, \exp\big((\mu-\mu_0)^2/2\sigma^2\big),
\end{equation}
where $1/\sigma^2$ is the second derivative
$\partial^2 f(\mu)/\partial \mu^2|_{\mu_0}=N_B {\rm sech}^2\mu_0$.
The integral converges because it runs from $-i\infty$ to $i\infty$
through the saddle point $\mu_0$ on the real axis.
Similarly, we have
\begin{equation}
\partial_E^2C_p(E)=\exp\big(f(\mu_0)\big)
\int {\rm d}\mu\,\mu^2 \,\exp\big((\mu-\mu_0)^2/2\sigma^2\big).
\end{equation}
Then the right-hand side of eq. (\ref{C_pm}) is
\begin{equation}
\frac{\displaystyle \int {\rm d}\mu\, \mu^2 \exp\big((\mu-\mu_0)^2/2\sigma^2\big)}
{\displaystyle \int {\rm d}\mu\,  \exp\big((\mu-\mu_0)^2/2\sigma^2\big)}-\mu_0^2
=(-\sigma^2+\mu_0^2)-\mu_0^2=-\sigma^2.
\end{equation}
It is therefore concluded that
\begin{equation}
-\frac{1}{CT^2}\le -\frac{1}{N_B{\rm sech}^2\mu_0},
\end{equation}
which leads to the bounds for the specific heat $0\le CT^2\le N_B{\rm sech}^2 \beta$
as is already known in the canonical ensemble.\cite{PTP81,Book2001}

Identities and inequalities for correlation functions can also be established.
Let us realize that the microcanonical delta constraint $\delta (E-H)$ with $E$ being
the control parameter plays a very similar role as the canonical Boltzmann factor
$\exp(-\beta H)$ with $\beta$ the control parameter.
It is then straightforward to apply the same argument as in the canonical case
to derive identities and inequalities for correlation functions.
If we take the example of the Gaussian disorder and use the notation $E_J\equiv -N_BJ_0$
(the NL conditions is $E=E_J$), the results are
\begin{eqnarray}
\left[ (\langle S_i S_j\rangle_E)^n \right]&=&
\left[ \langle S_i S_j\rangle_{E_J}(\langle S_i S_j\rangle_E)^n \right] \quad
(n=1,3,5,\cdots)
\label{I1}\\
\left[ P(m)\right]&=&\left[ P(q)\right]\quad (E=E_J)\label{I2}\\
\left|\left[ \langle S_i S_j\rangle_{E}\right]\right|
&\le&\left[\left| \langle S_i S_j\rangle_{E_J}\right|\right] \label{I3}\\
\left[ {\rm sgn}\langle S_i S_j\rangle_{E}\right]
&\le&\left[{\rm sgn} \langle S_i S_j\rangle_{E_J}\right] \label{I4},
\end{eqnarray}
where the suffix of the angular brackets specifies the value of the microcanonical energy.
The square brackets denote the configurational average.
The first relation (\ref{I1}) with $n=1$ shows that the ferromagnetic correlation
on the left-hand side is equal to the spin glass correlation on the right-hand side
if the system is on the NL.
The limit $|i-j|\to \infty$ yields $m=q$, where $m$ and $q$ are the ferromagnetic
and spin glass (Edwards-Anderson) order parameters, respectively.
The second identity (\ref{I2}) restates this fact from the perspective of the distribution
functions of the magnetization and the spin glass order parameter,
\begin{eqnarray}
P(m)&=&\frac{\sum_S \delta (Nm-\sum_i S_i)\delta (E-H)}{\sum_S \delta (E-H)}\\
P(q)&=&\frac{\sum_{S^{(1)},S^{(2)}} \delta (Nq-\sum_i S_i^{(1)}S_i^{(2)})
\delta (E-H^{(1)})\delta (E-H^{(2)})}{\sum_{S^{(1)},S^{(2)}} \delta (E-H^{(1)})\delta (E-H^{(2)})},
\end{eqnarray}
where $H^{(k)}$ is the Hamiltonian with the spins $S_i$ in eq. (\ref{Hamiltonian})
replaced by the spins $S_i^{(k)}$ of the $k (=1,2)$th replica.
This identity (\ref{I2}) proves that there is no replica symmetry breaking  on the NL
in the sense of non-trivial distribution $[P(q)]$ because the distribution
of magnetization $[P(m)]$ is always trivial.
The third relation (\ref{I3}) proves that the phase boundary between the ferromagnetic
and non-ferromagnetic phases below the multicritical point should be
either vertical or reentrant in the phase diagram.
The final inequality (\ref{I4}) implies that the number of mutually parallel
spin pairs takes its maximum value on the NL when we change the energy (and consequently
the temperature).
The system is thus in its most ordered state on the NL if we focus ourselves
to the spin orientation, ignoring the magnitude, as indicated by the signum function.
All these results are shared by the canonical ensemble.

The distribution of a local energy $-J_{12}S_1 S_2$ can be calculated similarly on the NL
using its gauge invariance.
The strategy is exactly the same as in the derivation of eq. (\ref{beta_Gaussian}).
The result is
\begin{equation}
 \left[\langle \delta (\epsilon +J_{12}S_1 S_2)\rangle_{E_J} \right]
 =c\, \exp\left(-\frac{N_B}{2(N_B-1)}(\epsilon -J_0)^2\right),
\end{equation}
where $c$ is the normalization constant.
The local energy naturally distributes in a Gaussian form around its mean $J_0$.

A different viewpoint can be introduced if we apply random fields and impose a microcanonical
constraint that the `staggered magnetization' along the random fields has a specific value.
For the Gaussian distribution of disorder, the entropy is
\begin{eqnarray}
S(E)&=&\frac{1}{C_J}\int {\rm D}J\,{\rm D}h\, \delta \big(N_BJ_0-\sum_{\langle ij\rangle} J_{ij}\big)
\delta \big(N h-\sum_i h_i\big)\nonumber\\
&&\cdot\log \sum_S \delta \big(E+\sum_{\langle ij\rangle} J_{ij}S_i S_j\big)
\delta \big(Nm-\sum_ih_i S_j\big).
\label{Sm}
\end{eqnarray}
The normalization is now
\begin{equation}
C_J=\int {\rm D}J\,{\rm D}h\, \delta \big(N_BJ_0-\sum_{\langle ij\rangle} J_{ij}\big)
\delta \big(Nh-\sum_i h_i\big),
\end{equation}
which can be evaluated as in eq. (\ref{C}) under the generalized NL condition
$E=-N_B J_0$ and $h=m$ to give
\begin{equation}
C_J(E,m)=\frac{1}{2\pi\sqrt{NN_B}}\,\exp\big(-E^2/(2N_B)-Nm^2/2\big).
\end{equation}
Then, when $E=-N_B J_0$ and $h=m$, the first and second derivatives of $S$ with respect to $m$ are
evaluated as before,
\begin{eqnarray}
\frac{\partial S}{\partial m}&=&\partial_m \log C_J(E,m)=-mN,\label{dSm}\\
\frac{\partial^2 S}{\partial m^2}&=
&\frac{\partial_m^2 C_J(E,m)}{C_J(E,m)}-\left(\frac{\partial_m C_J(E,m)}{C_J(E,m)}\right)^2
=-N.
\end{eqnarray}
Thus, for $m=h>0$, the entropy is a decreasing concave function of
the staggered magnetization on the NL ($E=-N_B J_0$).
In particular, for $m=h=0$, the vanishing value of the staggered magnetization
is thermodynamically stable in the sense that the entropy is maximum.
This is natural because the system is not in the spin glass phase ($q>0, m=0$)
on the NL ($q=m$) and thus cannot be staggered-magnetized along a given
unbiased random field (unbiased in the sense $h=0$).
For a finite value of $h (>0)$, the stable value of the staggered magnetization
is closer to zero than $m=h$ since the entropy will be larger for
smaller $m$ according to eq. (\ref{dSm}).

In summary, we have shown that all the results obtained from gauge symmetry in spin glasses
in the canonical ensemble can be reproduced in the microcanonical formulation of the same problem.
In particular, the microcanonical configurational average of the energy for the Gaussian
distribution agrees exactly with the corresponding canonical energy for any finite-size systems.
This implies complete ensemble equivalence for finite-size systems, an unusual phenomenon.
These results are valid for a generic system with arbitrary range of interactions
in arbitrary dimension including the infinite-range limit as long as the system is on the NL.
We have proved that there are no anomalies such as multiple values of the
temperature for a given energy or negative specific heat as observed in certain systems
with long-range interactions.\cite{Campa,Ant,Lynden,Thirring,spin1,spin2,spin3,spin4}
If ensemble inequivalence exists in spin glasses with long-range interactions,
it should happen away from the NL.
Preliminary calculations indeed suggest possible ensemble inequivalence away from the NL,
and the results will be reported in a forthcoming publication.

\end{document}